\documentclass[conference]{IEEEtran}
\IEEEoverridecommandlockouts
\usepackage{cite}
\usepackage{amsmath,amssymb,amsfonts}
\usepackage{algorithmic}
\usepackage{graphicx}
\usepackage{textcomp}
\usepackage{xcolor}
\usepackage{multicol,multirow}
\usepackage{xspace}
\usepackage{comment}
\newcommand{\ours}[0]{IMCorrect\xspace}
\def\BibTeX{{\rm B\kern-.05em{\sc i\kern-.025em b}\kern-.08em
    T\kern-.1667em\lower.7ex\hbox{E}\kern-.125emX}}
\begin{document}

\title{Recommendation Unlearning via Matrix Correction
%\\
%\thanks{This work was supported by the National Natural Science Foundation of China (NSFC) under Grants 61932007 and 62172106. Hansu Gu and Tun Lu are the corresponding authors.}
}

\author{
    \IEEEauthorblockN{Jiahao Liu\IEEEauthorrefmark{1}, Dongsheng Li\IEEEauthorrefmark{2}, Hansu Gu, Tun Lu\IEEEauthorrefmark{1}, Jiongran Wu\IEEEauthorrefmark{1}, Peng Zhang\IEEEauthorrefmark{1}, Li Shang\IEEEauthorrefmark{1} and Ning Gu\IEEEauthorrefmark{1}}
    \IEEEauthorblockA{\IEEEauthorrefmark{1} \textit{School of Computer Science, Fudan University,} Shanghai, China}
    \IEEEauthorblockA{\IEEEauthorrefmark{2} \textit{Microsoft Research Asia,} Shanghai, China}
    \IEEEauthorblockA{jiahaoliu21@m.fudan.edu.cn, dongsli@microsoft.com, hansug@acm.org, lutun@fudan.edu.cn\\ 23212010033@m.fudan.edu.cn, \{zhangpeng\_,lishang,ninggu\}@fudan.edu.cn}

}

\maketitle

\begin{abstract}
Recommender systems are important for providing personalized services to users, but the vast amount of collected user data has raised concerns about privacy (e.g., sensitive data), security (e.g., malicious data) and utility (e.g., toxic data). To address these challenges, recommendation unlearning has emerged as a promising approach, which allows specific data and models to be forgotten, mitigating the risks of sensitive/malicious/toxic user data. However, existing methods often struggle to balance completeness, utility, and efficiency, i.e., compromising one for the other, leading to suboptimal recommendation unlearning. In this paper, we propose an Interaction and Mapping Matrices Correction (\ours) method for recommendation unlearning. Firstly, we reveal that many collaborative filtering (CF) algorithms can be formulated as mapping-based approach, in which the recommendation results can be obtained by multiplying the user-item interaction matrix with a mapping matrix. Then, \ours can achieve efficient recommendation unlearning by correcting the interaction matrix and enhance the completeness and utility by correcting the mapping matrix, all without costly model retraining. Unlike existing methods, \ours is a whitebox model that offers greater flexibility in handling various recommendation unlearning scenarios. Additionally, it has the unique capability of incrementally learning from new data, which further enhances its practicality. We conducted comprehensive experiments to validate the effectiveness of \ours and the results demonstrate that \ours is superior in completeness, utility, and efficiency, and is applicable in many recommendation unlearning scenarios.
\end{abstract}

\begin{IEEEkeywords}
recommendation unlearning, recommender systems, collaborative filtering, machine unlearning
\end{IEEEkeywords}

\section{Introduction}
Recommender systems play a vital role in providing personalized services to users by learning their preferences through collected user data~\cite{ji2020dual}.
However, a major concern with these systems is the potential privacy breach since they retain the training data~\cite{carlini2019secret, fredrikson2015model, bourtoule2021machine}.
As a result, recommender systems may inadvertently leak user privacy~\cite{zhang2021membership, carlini2021extracting, zanella2020analyzing}.
To address this issue, users may want to delete specific interactions, preferences, or even all of their data, exercising their right to be forgotten as granted by recent regulations such as GDPR\footnote{https://gdpr-info.eu}, CCPA\footnote{https://oag.ca.gov/privacy/ccpa}, and PIPEDA\footnote{https://www.priv.gc.ca/en/privacy-topics/privacy-laws-in-canada/the-personal-information-protection-and-electronic-documents-act-pipeda}.
This necessitates the ability of recommender systems to forget the knowledge they have acquired, a process known as \textit{recommendation unlearning}~\cite{chen2022recommendation}.
Moreover, recommender systems are sensitive to data quality~\cite{schafer2007collaborative}, and the presence of malicious, toxic, out-of-date, or out-of-distribution data can hurt their performance~\cite{li2016data, jagielski2018manipulating, carlini2019secret}.
Removing such data can significantly improve the usability of the recommender system.

\begin{figure*}[htbp]
\centerline{\includegraphics[width=0.75\linewidth]{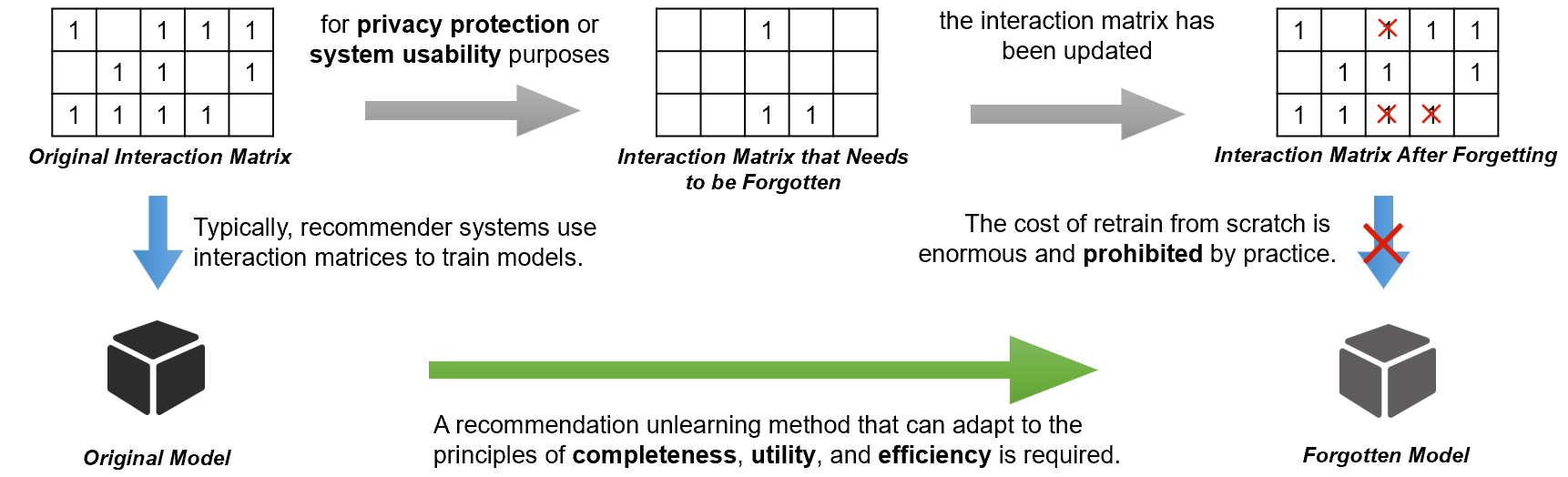}}
\caption{The process of recommendation unlearning.}
\label{fig:ru}
\end{figure*}

For recommendation unlearning to be effective, it should not only involve the deletion of collected user data but also the removal of the part of the model trained using that data.
The process should adhere to three fundamental principles~\cite{chen2022recommendation}:
\begin{itemize}
\item \textit{Completeness}. This principle requires completely eliminating the influence of the target data on the model, ensuring that no traces of the forgotten information remain.
\item \textit{Utility}. The accuracy and performance of the model after the unlearning process should be comparable to its performance before the forgetting process.
\item \textit{Efficiency}. The unlearning process should be efficient in terms of time and computational resources.
\end{itemize}
One straightforward method to achieve both completeness and utility is to retrain the model from scratch.
However, this approach is impractical due to its high training cost, making it inefficient for real-world applications.
Fig.~\ref{fig:ru} shows the process of recommendation unlearning.

To ensure the efficiency of forgetting, existing recommendation unlearning methods have made compromises between completeness or utility and efficiency.
RecEraser~\cite{chen2022recommendation} is based on the divide-aggregate architecture.
It first divides the data into several non-overlapping shards and then trains corresponding sub-models using data from each shard.
Finally, the results of each sub-model are aggregated.
When data needs to be forgotten, only the sub-model corresponding to the shard containing the target data and the aggregation layer need to be retrained.
On one hand, by avoiding the need to retrain the model on the entire dataset, the efficiency of forgetting is improved.
On the other hand, since dividing data into shards inevitably leads to the loss of collaborative information, the utility of the model is compromised.
This type of method, which improves the retrain efficiency while ensuring complete forgetting, is known as \textit{exact recommendation unlearning}.
Several other methods~\cite{liu2022forgetting,li2023selective,zhang2023recommendation} achieve unlearning using fine-tuning without changing the original workflow.
When data needs to be forgotten, they use influence functions to quantify the impact of the target data on the model and further eliminate this influence.
Due to the need of reducing the number of updated parameters in such methods to make the calculation of the Hessian matrix feasible, it may raise concerns about completeness.
This type of method, which avoids retraining, and sacrifices the completeness of forgetting while improving efficiency, is called \textit{approximate recommendation unlearning}.

Collaborative Filtering (CF) has been widely used to provide personalized recommendations to users by modeling their interactions with items~\cite{hu2008collaborative,koren2009matrix,rendle2012bpr}.
In this paper, we focus on the form of obtaining prediction matrices in CF models and classify them into two categories: \textit{embedding-based CF methods} and \textit{mapping-based CF methods}.
In the embedding-based CF methods, embeddings for users and items are first obtained, and then scores for users and items are often calculated based on the similarity between these embeddings to generate the prediction matrix, e.g., dot product.
On the other hand, mapping-based CF methods directly map the interaction matrix to the prediction matrix via a mapping matrix.
We conduct a thorough analysis of the mapping-based CF methods and reveal that the mapping matrix holds the essence of the similarity matrix between items, encompassing global structural information.
Moreover, we show that the embedding-based CF methods can be naturally formulated as mapping-based CF methods, decoupling user interaction data from item similarity.

Based on the mapping matrix formulation, we propose a model-agnostic recommendation unlearning method named \textbf{\ours} (\textbf{I}nteraction and \textbf{M}apping {M}atrices \textbf{Correct}ion).
%For mapping-based CF methods, 
\ours achieves highly efficient forgetting by correcting only the interaction matrix without the need for costly retraining, while preserving the model's performance.
Additionally, we analyze how the target data influences the mapping matrix and subsequently correct it to eliminate this influence, enhancing the completeness and utility of unlearning.
The whitebox nature of \ours makes it a flexible approach compared to existing methods, enabling it to address various recommendation unlearning scenarios effectively.

Specifically, we analyze \ours in three different application scenarios: 1) handling out-of-distribution data, 2) handling out-of-date data, and 3) addressing attack data.
We also explore how \ours can be applied to recommendation unlearning while adhering to privacy protection requirements.
A noteworthy advantage of \ours is its capability to not only forget existing interaction data (decremental learning) but also learn from new interaction data (incremental learning), accommodating recommender systems that continuously update their models due to evolving user preferences~\cite{wang2020toward,liu2022parameter,liu2023triple,xia2022fire}.

We instantiate \ours on SLIM~\cite{ning2011slim}, GF-CF~\cite{shen2021powerful}, and MF~\cite{hu2008collaborative} and demonstrate its superiority through comprehensive experiments.
Firstly, we validate the general capabilities (completeness, utility, and efficiency) of \ours under three types of attacks.
Then, we perform case studies on recommendation unlearning tasks in various real-world scenarios to demonstrate the abilities of \ours in different situations.
Experimental results show that \ours achieves outstanding performance in completeness, utility, and efficiency, while effectively handling real-world unlearning tasks.

Our main contributions can be summarized as follows:
\begin{itemize}
\item By analyzing the mapping matrix and its implications in CF methods, we shed light on the inner workings of mapping-based CF methods and their connections to embedding-based CF methods.
\item We proposed a new model-agnostic method for approximate recommendation unlearning, \ours, which is a whitebox model that offers greater flexibility, making it easily applicable to various recommendation scenarios.
\item We instantiated \ours on various CF methods and conducted comprehensive experiments to validate its superiority in terms of general capabilities and scenario-specific abilities, showcasing its practical value in real-world applications.
\end{itemize}

\section{Related Work}
In this section, we introduce the related works on collaborative filtering and recommendation unlearning.

\subsection{Collaborative Filtering}
Collaborative filtering is a kind of widely used recommendation algorithm that extracts users' preferences and item attributes from user-item interaction data to provide personalized recommendations~\cite{koren2021advances, liu2023personalized, chen2020efficient}.
Recommender systems effectively alleviate the problem of information overload, and collaborative filtering algorithms have continuously evolved since their inception.

User-based Collaborative Filtering (User-CF)\cite{herlocker1999algorithmic} and Item-based Collaborative Filtering (Item-CF)\cite{sarwar2001item} calculate Pearson correlation coefficients between users and items based on their interaction records to generate recommendations.
To address the sparsity problem caused by incomplete data, Matrix Factorization (MF) techniques have been proposed and applied in collaborative filtering~\cite{koren2008factorization, koren2009matrix, hu2008collaborative}.
MF models user interests and item attributes as latent vectors in the same latent space, composed of several latent factors, and measures the interaction scores between users and items through dot products.
The latent vectors can be obtained through Singular Value Decomposition (SVD)\cite{sarwar2000application}, reconstruction loss optimization\cite{koren2009matrix}, or Bayesian Personalized Ranking (BPR)\cite{rendle2012bpr}.
SLIM~\cite{ning2011slim} and EASE\^{}R~\cite{steck2019embarrassingly} efficiently and accurately provide top-$K$ recommendations through sparse linear methods.

In recent years, collaborative filtering algorithms combined with deep learning have become very popular.
AutoRec~\cite{sedhain2015autorec} predicts users' interests using an autoencoder architecture, where the encoder and decoder are implemented through multi-layer perceptrons (MLP).
NCF~\cite{he2017neural} introduces non-linear mechanisms in neural networks to enhance the expressive power of recommendation models.
NGCF~\cite{wang2019neural} further considers multi-hop information in the user-item interaction graph using graph neural networks.
LightGCN~\cite{he2020lightgcn} removes some unnecessary modules from NGCF to achieve more efficient and accurate predictions.

\subsection{Recommendation Unlearning}
\begin{figure}[b]
\centerline{\includegraphics[width=0.75\linewidth]{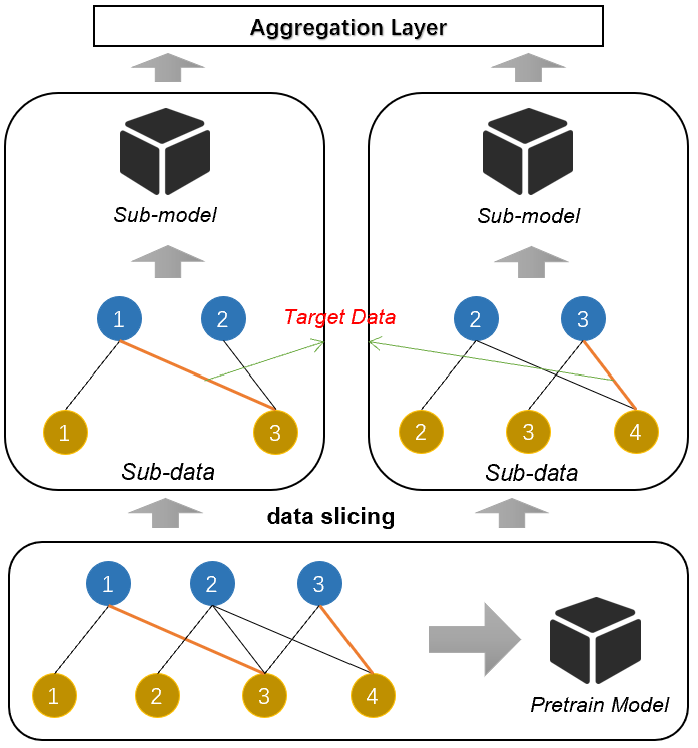}}
\caption{The process of RecEraser.}
\label{fig:rec}
\end{figure}

Machine unlearning~\cite{bourtoule2021machine}, also known as selective forgetting~\cite{golatkar2020eternal} or data removal/deletion~\cite{guo2019certified}, refers to the process of removing the influence of specific data from a trained model.
Existing machine unlearning works often focus on tasks in computer vision or natural language processing~\cite{schelter2021hedgecut,baumhauer2022machine,bourtoule2021machine}.
Due to architectural differences, these methods cannot be directly applied to recommendation tasks~\cite{liu2022forgetting}.
Unlike differential privacy~\cite{dwork2014algorithmic,gao2020dplcf,dwork2006differential}, which focuses on protecting rather than deleting private information, recommendation unlearning requires the recommender system to be able to delete user data and its associated model.
According to the degree of achievement regarding the completeness principle, recommendation unlearning can be divided into exact recommendation unlearning and approximate recommendation unlearning.

\subsubsection{Exact Recommendation Unlearning}
Exact unlearning refers to methods that can guarantee complete forgetting of data (completeness)~\cite{chen2022graph,he2021deepobliviate,bourtoule2021machine}.
These methods often use a divide-aggregate architecture to improve retraining efficiency by partitioning the model~\cite{ginart2019making,schelter2021hedgecut} and dividing the data~\cite{bourtoule2021machine,cao2015towards,yan2022arcane}.
SISA (Sharded, Isolated, Sliced, and Aggregated)~\cite{bourtoule2021machine} reduces the amount of data involved in retraining by dividing the data into several disjoint shards.
For graph data, GraphEraser~\cite{bourtoule2021machine} uses a balanced graph partition method during sharding and employs a learning-based method for static weighting during aggregation.

Following the concepts of SISA and GraphEraser, RecEraser~\cite{chen2022recommendation} introduces three novel data partition algorithms to divide the training data into balanced groups, ensuring that similar users and/or items are grouped together and protects collaborative information to a certain extent, which is crucial for collaborative filtering algorithms.
RecEraser then uses sub-models to train each group separately.
Finally, considering that different subgroups contribute differently to different user-item pairs, RecEraser proposes an adaptive attention-based aggregation method to further enhance model utility.
When data needs to be forgotten, only the sub-model corresponding to the shard containing the target data needs to be retrained.
Since it involves retraining, completeness can be ensured.
LASER~\cite{li2022making} is concurrent work with RecEraser, following a similar approach but with different specific details.

\textbf{We acknowledge the contributions of RecEraser, but there are some shortcomings that need to be addressed.}
For ease of description, we have illustrated the process of data slicing, sub-model training, aggregation, and forgetting in RecEraser, as shown in Fig.~\ref{fig:rec}.
\begin{itemize}
\item Collaborative filtering, as an association-sensitive task, contains a significant amount of collaborative information in the data~\cite{shi2014collaborative}.
Although RecEraser tries to preserve collaborative information as much as possible through carefully designed sharding and aggregation, the partitioning operation inevitably disrupts collaborative information, leading to a degradation of model utility.
As shown in Fig.~\ref{fig:rec}, interactions of the same user or item may be stored in different shards, making it challenging for sub-models to fully capture user interaction patterns.
\item RecEraser can only handle individual forgetting requests, and when requests arrive in batch form, the efficiency gained from sharding for forgetting may diminish.
As shown in Fig.~\ref{fig:rec}, when multiple target data need to be forgotten simultaneously, the system may be required to retrain multiple sub-models simultaneously.
\item RecEraser does not completely avoid retraining, which reduces its efficiency.
As shown in Fig.~\ref{fig:rec}, both the pretrain phase and the aggregation phase require training with all the data, and each forgetting process requires retraining the sub-model and the aggregation layer.
\item The divide-aggregate architecture on which RecEraser relies has its own drawbacks:
1) it disrupts the original workflow and is difficult to deploy;
2) we found in experiments that the performance of sub-models varies significantly, resulting in high variance;
3) as shown in Fig.~\ref{fig:rec}, the architecture may store multiple sets of parameters for the same user or item, leading to a substantial increase in parameter size.
\end{itemize}

\subsubsection{Approximate Recommendation Unlearning}
Due to the aforementioned shortcomings, approximate recommendation unlearning has become a hot topic in current research.
Approximate unlearning refers to methods that can sacrifice some completeness to improve the efficiency of forgetting and the utility of the model, achieving only statistical forgetting~\cite{izzo2021approximate,neel2021descent,golatkar2020forgetting}.
These methods often directly eliminate the influence of target data on the model through inverse gradient-based update strategies to achieve forgetting without retraining.
The tool used to measure the influence of target data on a trained model is called the influence function, which is a classic concept originating from robust statistics~\cite{hampel1974influence,koh2017understanding,koh2019accuracy}.
Since calculating the Hessian matrix is required, existing work that applies influence functions for machine unlearning needs to optimize for efficiency~\cite{wu2022puma,mehta2022deep,wu2020deltagrad}.
The influence function has been widely used in recommendation tasks, for example, some works use influence functions to achieve recommendation attacks~\cite{yi2014robust,fang2020influence,wu2021triple,zhang2020practical}, recommendation explanations~\cite{cheng2019incorporating}, and recommendation debiasing~\cite{yu2020influence}.

Existing recommendation unlearning works have adapted the use of influence functions from classification or regression tasks~\cite{guo2019certified,golatkar2020eternal,mahadevan2021certifiable,sekhari2021remember} to the collaborative filtering task.
Their approaches involve utilizing influence functions with 2nd-order optimization methods, such as Newton and quasi-Newton methods, which are advantageous due to their fast final convergence, to accelerate the finetuning process~\cite{tsai2014incremental}.
The differences between these works mainly lie in how they optimize to speed up the computation of the Hessian matrix.
AltEraser~\cite{liu2022forgetting} decomposes the large optimization problem into several smaller independent sub-problems using alternating optimization~\cite{bezdek2003convergence,zhou2008large}.
Alternating optimization allows each sub-problem to be solvable and they can be solved in parallel, thus improving optimization efficiency.
In the specific implementation, AltEraser also utilizes optimization techniques like Hessian-free Newton~\cite{martens2010deep}, Ad Hoc Newton~\cite{mahadevan2021certifiable}, and others.
SCIF~\cite{li2023selective} selectively updates user embeddings to reduce the number of parameters that need to be updated and preserves the collaborative information across the remaining users and items to enhance the model's utility.
SCIF computes the Hessian-Vector Product following~\cite{basu2020second} and~\cite{koh2017understanding} to reduce computational overhead. 
IFRU~\cite{zhang2023recommendation} extends the influence function to quantify the influence of unusable data in the computational graph aspect and proposes an importance-based pruning algorithm to reduce the cost of the influence function.
Yuan et al.\cite{yuan2023federated} proposed an unlearning method for federated recommendation, which removes a user's contribution by rolling back and introduces a small-sized negative sampling method to reduce the number of item embedding updates, along with an importance-based update selection mechanism to store only important model updates.
Unlearn-ALS~\cite{xu2023netflix} modifies the fine-tuning procedure under Alternating Least Squares optimization for bi-linear recommendation models.
Approximate recommendation unlearning cannot guarantee completeness, so they use membership inference~\cite{shokri2017membership,yu2021does,wu2020characterizing,salem2018ml,zhang2021membership}, a binary classification problem that determines whether the target data appears in the training set of the model after forgetting, to check the completeness of forgetting.

\begin{figure}[t]
\centerline{\includegraphics[width=0.95\linewidth]{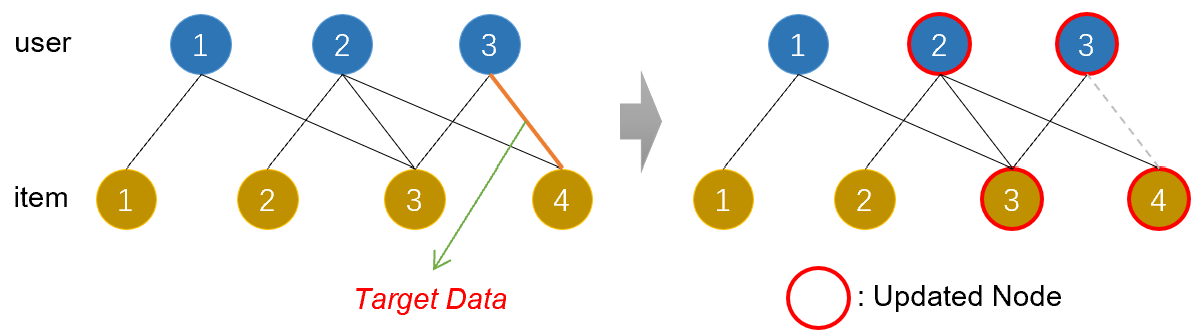}}
\caption{The recommendation unlearning method based on the influence function only updates the target data and its neighbors.}
\label{fig:if}
\end{figure}

Due to the inability of approximate recommendation unlearning methods to efficiently compute the Hessian matrix, they each reduce the number of parameters that need to be updated in different ways.
While this approach can improve the efficiency of forgetting, it also means that \textbf{these methods are unable to completely eliminate the influence of target data}, especially in non-convex models like deep neural networks.
As shown in Fig.~\ref{fig:if}, to reduce the number of updated parameters, these methods only update the embeddings of the target data and its limited neighbors, which results in an incomplete eradication of the influence of target data on the model.

\section{Analysis}
\label{sec:analysis}
In this section, we first introduce four representative collaborative filtering methods.
Then, we analyze the essence of the mapping matrix in mapping-based CF methods.
Finally, we explain how to compute the mapping matrix for embedding-based CF methods to formulate them as mapping-based CF methods.
Assume that there are $m$ users and $n$ items, and the interaction matrix $R\in\{0,1\}^{m\times n}$ indicates the interactions between users and items, where the value $1$ represents an interaction, and $0$ represents no interaction.
We have listed some important symbols and their definitions in the TABLE~\ref{tab:sym}.

\begin{table}[htbp]
\caption{The Notations Used in This Paper}
\begin{center}
\begin{tabular}{c|c}
\hline
Symbol & Definition                                              \\ \hline
$R$      & original interaction matrix                             \\
$W$      & original mapping matrix                                          \\
$\bar{R}$   & interaction matrix that needs to be forgotten           \\
$\bar{W}$   & mapping matrix that needs to be forgotten           \\
$\tilde{R}$ & interaction matrix after forgetting                           \\
$\tilde{W}$ & mapping matrix after forgetting                               \\
$\hat{R}$   & predicted interaction matrix                            \\
$P$      & user embedding matrix                                         \\
$Q$      & item embedding matrix                                         \\
$\mathbf{a}_{i \cdot}$ & the $i$-th row of matrix $A$                                 \\
$\mathbf{a}_{\cdot j}$ & the $j$-th column of matrix $A$                              \\
$a_{ij}$    & the element of the $i$-th row and $j$-th column of matrix $A$ \\ \hline
\end{tabular}
\label{tab:sym}
\end{center}
\end{table}

\subsection{Details of SLIM, GF-CF, AutoRec and MF}\label{sec:fga9}
In this section, we first introduce four representative collaborative filtering methods: SLIM~\cite{ning2011slim}, GF-CF~\cite{shen2021powerful}, AutoRec~\cite{sedhain2015autorec}, and MF~\cite{sarwar2000application,hu2008collaborative}.
For simplicity of exposition, we omit regularization terms for all methods.

1) SLIM learns a mapping matrix $W_{SLIM}\in\mathbb{R}^{n\times n}$ through the following optimization problem:
\begin{equation}\label{eqn:1}
\begin{aligned}
W_{SLIM}=\mathop{\arg\min}_W\|R-RW\|_2\text{,}\\ \text{s.t. }W\ge0\text{ and }diag(W)=0\text{.}
\end{aligned}
\end{equation}
Then, the prediction matrix is computed as follows:
\begin{equation}
\hat{R}_{SLIM}=RW_{SLIM}\text{.}
\end{equation}

2) GF-CF first performs truncated singular value decomposition (SVD) on $R$:
\begin{equation}\label{eq:iba0}
U\text{, }\Sigma\text{, }V=svd_k(R)\text{,}
\end{equation}
where $U\in\mathbb{R}^{m\times k}$, $\Sigma\in\mathbb{R}^{k\times k}$, and $V\in\mathbb{R}^{n\times k}$.
Then, it constructs a low-pass graph filter $W_{GF-CF}\in\mathbb{R}^{n\times n}$ as follows: 
\begin{equation}
W_{GF-CF}=VV^\top\text{.}
\end{equation}
Finally, the prediction matrix is computed as follows:
\begin{equation}
\hat{R}_{GF-CF}=RW_{GF-CF}=RVV^\top\text{.}
\end{equation}
Here, $R$ multiplied by $V$ in the first step represents mapping the interaction matrix to the Fourier space and keeping only the low-frequency components (Fourier transformation), and the subsequent multiplication by $V^\top$ represents mapping back the low-frequency signals in the Fourier space to the original space (inverse Fourier transformation).

3) AutoRec utilizes an autoencoder $W_{AutoRec}:\mathbb{R}^{z^+\times n}\rightarrow\mathbb{R}^{z^+\times n}$ to compress and reconstruct $R$:
\begin{equation}
W_{AutoRec}(R)=d(e(R))\text{,}
\end{equation}
where $e:\mathbb{R}^{z^+\times n}\rightarrow\mathbb{R}^{z^+\times k}$ and $d:\mathbb{R}^{z^+\times k}\rightarrow\mathbb{R}^{z^+\times n}$ are encoding and decoding functions realized by multi-layer perceptrons (MLPs), respectively, and $z^+$ represents any positive integer.
By minimizing the reconstruction loss, the optimal autoencoder $W_{AutoRec}^*$ is obtained:
\begin{equation}\label{eq:ica8}
W_{AutoRec}^*=\mathop{\arg\min}_{W_{AutoRec}}\|R-W_{AutoRec}(R)\|_2\text{.}
\end{equation}
Finally, the prediction matrix is computed as follows:
\begin{equation}
\hat{R}_{AutoRec}=W_{AutoRec}^*(R)\text{.}
\end{equation}

4) Matrix factorization (MF) can be implemented in various ways. After obtaining $U$, $\Sigma$, and $V$ through truncated SVD {using~\eqref{eq:iba0}}, Sarwar et al.~\cite{sarwar2000application} computed user embeddings $P_{SVD}$ and item embeddings $Q_{SVD}$ as follows:
\begin{equation}
P_{SVD}=U\Sigma^{1/2}\text{, }Q_{SVD}=V\Sigma^{1/2}\text{.}
\end{equation}
Then, the prediction matrix is computed as follows: 
\begin{equation}
\hat{R}_{SVD}=P_{SVD}Q_{SVD}^\top
\end{equation}
Hu et al.~\cite{hu2008collaborative}, on the other hand, obtain user embeddings $P_{Re}\in\mathbb{R}^{m\times k}$ and item embeddings $Q_{Re}\in\mathbb{R}^{n\times k}$ by minimizing the reconstruction loss as follows:
\begin{equation}
P_{Re}\text{, }Q_{Re}=\mathop{\arg\min}_{P,Q}\|R-PQ^\top\|_2\text{.}
\end{equation}
Then, the prediction matrix is computed as follows:
\begin{equation}
\hat{R}_{Re}=P_{Re}Q_{Re}^\top\text{.}
\end{equation}
In fact, $\hat{R}_{SVD}$ represents the best rank-$k$ approximation of $R$ under the squared reconstruction loss.

We have introduced classic and representative methods, while many recent works can be seen as extensions of them.
For example, NCF~\cite{he2017neural} and LightGCN~\cite{he2020lightgcn} can be viewed as improvements to MF, differing only in the way they obtain user embeddings and item embeddings.
However, recent research has shown that some of these newly proposed methods do not always outperform these classic methods, and mapping-based CF methods and embedding-based CF methods each have their advantages and disadvantages on different datasets~\cite{zhu2022bars}.

\begin{figure}[t]
\centerline{\includegraphics[width=0.85\linewidth]{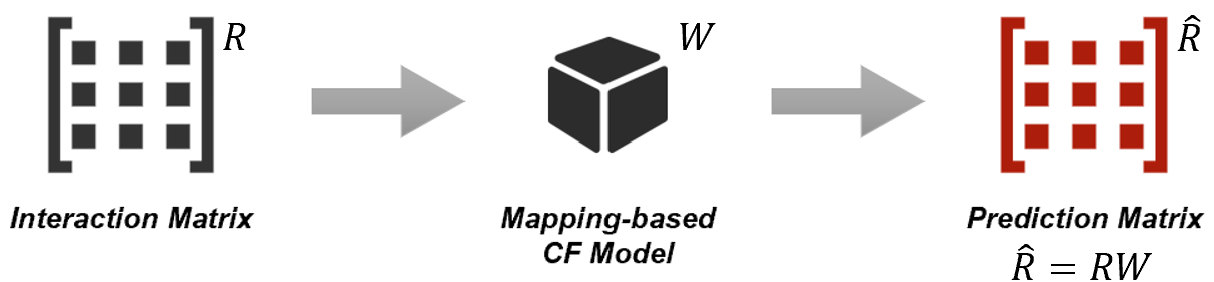}}
\caption{The process of obtaining the prediction matrix by mapping-based CF methods.}
\label{fig:map}
\end{figure}

\subsection{Mapping Matrix of Mapping-based CF Methods}\label{sec:ddf8}
As shown in Fig.~\ref{fig:map}, for SLIM, GF-CF, or AutoRec, they can all be unified in the form as the result of multiplying the interaction matrix $R$ with a mapping matrix $W\in\mathbb{R}^{n\times n}$, to yield the prediction matrix:
\begin{equation}\label{eq:da9s}
\hat{R}=RW\text{.}
\end{equation}
In the case of AutoRec, its encoder and decoder use MLPs to approximate the Fourier transform and inverse transform processes in GF-CF.

Equation~\eqref{eq:da9s} shows that each column of $\hat{R}$ is a linear combination of all columns of $R$, where the combination coefficients are given by the corresponding column of $W$.
For example, for the $j$-\text{th} column of $\hat{R}$, we have:
\begin{equation}
\hat{\mathbf{r}}_{\cdot j}=\sum_{i=1}^nw_{ij}\mathbf{r}_{\cdot i}\text{.}
\end{equation}
Essentially, all mapping-based CF methods minimize the reconstruction loss, aiming to make $R$ and $RW$ to be as close as possible.
In the case of GF-CF, $W$ is obtained through numerical methods, while SLIM and AutoRec optimize the reconstruction loss using~\eqref{eqn:1} and~\eqref{eq:ica8} to obtain $W$ respectively.
It is evident that the reconstruction loss is minimized when $W$ is an identity matrix.
To avoid finding trivial solutions, SLIM imposes a constraint that the diagonal elements of $W$ are set to zero, meaning that a column cannot participate in its own reconstruction.
On the other hand, AutoRec and GF-CF impose rank constraints on $W$.

Generally, the more similar (i.e., higher co-occurrence frequency) the interaction patterns between two items, the larger the corresponding weight (combination coefficient) should be when reconstructing through linear combinations.
Since the reconstruction loss is related to the interaction patterns of items and is a global measure, \textbf{the essence of $W$ reflects the similarity between items with the ability to capture global structural information}.
Additionally, the spectral domain operations in GF-CF enable the consideration of global structural information in the spatial domain, further supporting the interpretation of $W$ as a measure of item similarity with global structural information.

\begin{figure}[t]
\centerline{\includegraphics[width=0.85\linewidth]{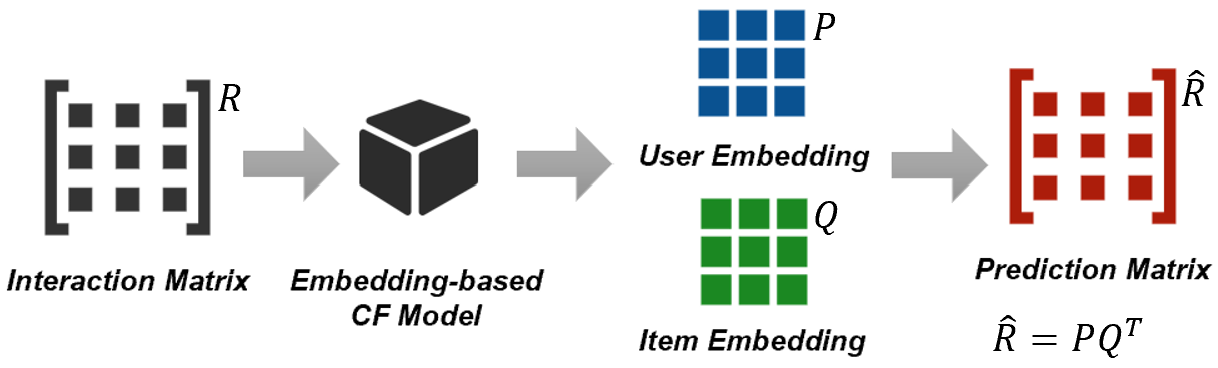}}
\caption{The process of obtaining the prediction matrix by embedding-based CF methods.}
\label{fig:emb}
\end{figure}
\subsection{Mapping Matrix in Embedding-based CF Methods}
After recognizing that the essence of the mapping matrix is a measure of item similarity that considers global structural information, we can compute it from the embedding-based CF methods.
Specifically, as shown in Fig.~\ref{fig:emb}, embedding-based CF methods first obtain user embeddings $P$ and item embeddings $Q$, which also contain global structural information.
This means that users and items with similar interaction patterns have similar embeddings.
We can compute the mapping matrix as follows:
\begin{equation}
W=QQ^T\text{,}
\end{equation}
and then use{~\eqref{eq:da9s}} to calculate the prediction matrix $\hat{R}$.

\textbf{This implies that embedding-based CF methods can be naturally transformed into mapping-based CF methods}.
Although the resulting prediction matrix may differ from $PQ^\top$ to some extent, this transformation has significant advantages for recommendation unlearning.
In mapping-based CF methods, the personal interaction records of users (sensitive information) are decoupled from the item similarity information.
On the other hand, in the embedding-based CF methods, user embeddings and item embeddings are semantically coupled together, which can cause trouble for recommendation unlearning.

\section{Method}
In this section, we first describe the recommendation unlearning problem, then introduce our proposed method, \ours, and finally explain how \ours can be applied in different scenarios.

\begin{figure*}[htbp]
\centerline{\includegraphics[width=0.75\linewidth]{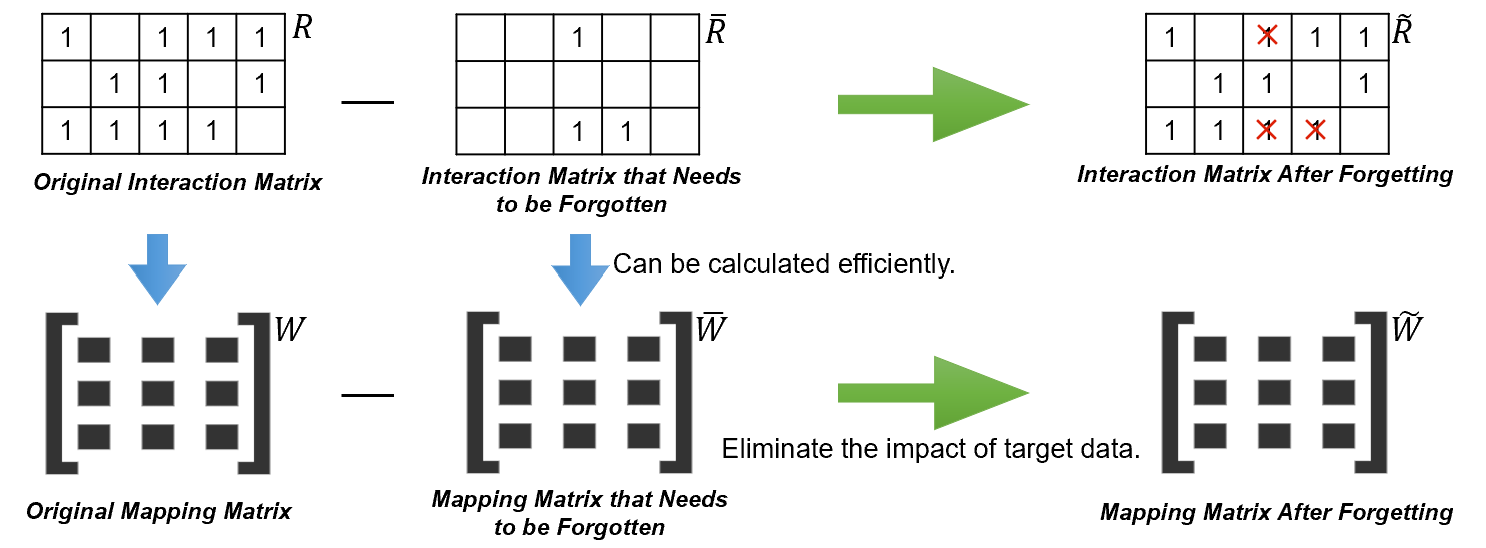}}
\caption{The unlearning process of \ours.}
\label{fig:ours}
\end{figure*}
\subsection{Problem Description}
Assuming that we have obtained the original mapping matrix $W$ based on the original interaction matrix $R$ using the methods described in Section \ref{sec:analysis} and have obtained the original prediction matrix $\hat{R}$ according to{~\eqref{eq:da9s}}.
Now, a small portion of elements in $R$, denoted as $\bar{R}\in\{0,1\}^{m\times n}$, has been flipped to create a new interaction matrix $\tilde{R}\in\{0,1\}^{m\times n}$.
The problem is how to quickly obtain the new prediction matrix $\hat{\tilde{R}}$.

Since \ours has the ability to both forget existing interaction data (decremental learning) and learn new interaction data (incremental learning), the flipping can be either a negative flip (from 1 to 0) or a positive flip (from 0 to 1).
We assume that the interactions in $\bar{R}$ are either all part of positive flips or all part of negative flips.

If an element is updated in the negative flips, it means that its corresponding interaction needs to be unlearned, which refers to recommendation unlearning.
In this case, we have:
\begin{equation}
\tilde{R}=R-\bar{R}\text{.}
\end{equation}

If an element is updated in the positive flips, it means that its corresponding interaction needs to be learned, which refers to incremental recommendation.
In this case, we have:
\begin{equation}
\tilde{R}=R+\bar{R}\text{.}
\end{equation}
This paper mainly focuses on the performance of \ours in the recommendation unlearning task, but \ours can be easily extended to incremental learning.

\subsection{\ours}
In \ours, we can achieve recommendation unlearning by correcting the interaction matrix and/or the mapping matrix.

\subsubsection{Correction of Interaction Matrix}
In Section~\ref{sec:ddf8}, we explained the meaning of the mapping matrix $W$ in mapping-based CF methods from a reconstruction perspective during the learning phase.
In the inference phase, after learning the mapping matrix $W$, each row of the prediction matrix in~\eqref{eq:da9s} can be considered as the result of aggregating a user's interaction history using $W$, where the $u$-\text{th} row of $\hat{R}$ is:
\begin{equation}\label{eq:o9an}
\begin{aligned}
\hat{\mathbf{r}}_{u\cdot}=\mathbf{r}_{u\cdot}W&=\mathbf{r}_{u\cdot}(\mathbf{w}_{\cdot 1},...,\mathbf{w}_{\cdot n})\\&=(\mathbf{r}_{u\cdot}\cdot\mathbf{w}_{\cdot 1},...,\mathbf{r}_{u\cdot}\cdot\mathbf{w}_{\cdot n})\\&=(\Sigma_{i=1}^{n}r_{ui}w_{i1},...,\Sigma_{i=1}^{n}r_{ui}w_{in})\text{.}
\end{aligned}
\end{equation}
\textbf{This implies that mapping-based CF methods essentially use a matrix $W$, which contains global structural information of item-item similarities, to aggregate the interaction matrix and achieve smoothing of the original interaction matrix $R$, resulting in the prediction matrix $\hat{R}$.}

In~\eqref{eq:o9an}, if a user $u$ has interacted with item $i$ ($r_{ui}=1$), then item $i$ will contribute to the aggregation score for user $u$, with the weights being the similarity between item $i$ and the corresponding items.
On the other hand, if a user $u$ has not interacted with item $i$ ($r_{ui}=0$), then item $i$ will not contribute to the aggregation score for user $u$.
This indicates that, for a trained mapping matrix, users' interaction history (the rows in the interaction matrix) completely determines their recommendation results.
Therefore, we can correct the interaction matrix to achieve recommendation unlearning (or incremental recommendation):
\begin{equation}\label{eq:gn9m}
\hat{\tilde{R}}=\tilde{R}W\text{.}
\end{equation}
\textbf{This means that with no additional training cost, we can achieve recommendation unlearning (or incremental recommendation).}

\subsubsection{Correction of Mapping Matrix}
The slight changes in the interaction matrix will obviously lead to slight changes in the mapping matrix $W$.
While~\eqref{eq:gn9m} is already capable of handling some recommendation unlearning tasks, in order to correct the prediction matrix more accurately and thoroughly, it is necessary to calculate a new mapping matrix $\tilde{W}$ based on the new interaction matrix $\tilde{R}$.
However, instead of obtaining a new mapping matrix $\tilde{W}$ through retraining as described in Section~\ref{sec:analysis}, which is highly inefficient, we achieve this by correcting the original mapping matrix $W$.

% The desired new mapping matrix $\tilde{W}$ should be learned from the new interaction matrix $\tilde{R}$ using the method described in Section~\ref{sec:analysis}.
The original mapping matrix $W$ is learned from the original interaction matrix $R$ using the method described in Section~\ref{sec:analysis}, which considers both the new interaction matrix $\tilde{R}$ and interaction matrix $\bar{R}$ that needs to be forgotten.
As we have recognized, the essence of the mapping matrix is a similarity matrix, and in the collaborative filtering scenario, item-item similarities are determined by the topological relationships in the bipartite graph composed of users and items.
Therefore, \textbf{the essence of the problem in correcting the mapping matrix is how to quantify the impact of updates in the link relationships between nodes in the bipartite graph on item-item similarities and eliminate this impact}.

Taking negative flip as an example, it refers to the process of forgetting $\bar{R}$, the interactions that need to be updated from the original interaction matrix $R$, resulting in a new interaction matrix $\tilde{R}$.

On one hand, this process reduces the co-occurrence frequency between items in $\bar{R}$ and other items, leading to a decrease in their similarity.
The extent of reduction depends on the number of times an item appears in $\bar{R}$, and we represent this effect as follows:
\begin{equation}\label{eq:dba9}
W'=(\frac{c_1-\bar{c}_1}{c_1}\mathbf{w}_{\cdot 1},...,\frac{c_n-\bar{c}_n}{c_n}\mathbf{w}_{\cdot n})\text{,} 
\end{equation}
where $c_i=\Sigma_{u=1}^mr_{ui}$ and $\bar{c}_i=\Sigma_{u=1}^m\bar{r}_{ui}$ are the number of interactions of item $i$ in the original interaction matrix $R$ and the interaction matrix $\bar{R}$ that needs to be forgotten, respectively.
For positive flip, a similar conclusion can be drawn, where the newly added interactions in $\bar{R}$ increase the similarity between items, resulting in
\begin{equation}\label{eq:bmq0}
W'=(\frac{c_1+\bar{c}_1}{c_1}\mathbf{w}_{\cdot 1},...,\frac{c_n+\bar{c}_n}{c_n}\mathbf{w}_{\cdot n})\text{.}
\end{equation}
Equations~\eqref{eq:dba9} and~\eqref{eq:bmq0} indicate that the more times an item appears in $\bar{R}$, the greater its impact on the similarity with other items.

\begin{table*}[t]
\caption{Three Application Scenarios of \ours for Improving System Usability}
\begin{center}
\begin{tabular}{l|p{4.75cm}|p{4.75cm}|p{4.75cm}}
\hline
\textbf{Scenario}       & Out-of-distribution data & Out-of-date data & Attack data \\ \hline
\textbf{Explanation} & Interactions outside a user's true interests. 
& Interactions caused by outdated short-term user interests. 
& Malicious interactions initiated by users. \\ \hline
\textbf{Case}        & A man who is not interested in female products buys a lipstick as a birthday gift for his girlfriend on an e-commerce platform. As a result, the e-commerce platform keeps recommending products related to lipsticks based on this purchase record. In this case, the man may want the system to forget this lipstick purchase interaction. & A user has already purchased a smartphone on an e-commerce platform and no longer intends to make any more purchases. However, the recommender system continues to suggest smartphones to this user based on their recent search and browsing history. In this scenario, the system should initiate forgetting of the user's recent behavior. & On a short video platform, some authors may generate fake views to make their videos appear popular, tricking the recommender system into promoting these videos to more users. Once the system identifies such interactions, it should promptly forget them to reduce the likelihood of promoting such videos to other users. \\ \hline
\textbf{Usage}       & Both the interaction matrix and the mapping matrix need to be updated.                       & Only the interaction matrix needs to be updated. 
& Only the mapping matrix needs to be updated. \\ \hline
\end{tabular}
\label{tab:case}
\end{center}
\end{table*}

On the other hand, for negative flip, since the interactions in $\bar{R}$ are removed, the pairs of items that have high similarity due to similar interaction patterns in $\bar{R}$ should have lower similarity in $\tilde{W}$ than in $W'$.
For positive flip, the pairs of items that have high similarity in $\bar{R}$ due to similar interaction patterns should have higher similarity in $\tilde{W}$ than in $W'$, as the interactions in $\bar{R}$ are added to $R$.
Since the number of interactions in $\bar{R}$ is much lower than that in $R$ and $\tilde{R}$, the method introduced in Section~\ref{sec:analysis} can quickly learn the mapping matrix $\bar{W}$ corresponding to $\bar{R}$.
After obtaining $\bar{W}$, we perform column-wise weighting on it as follows:
\begin{equation}
\bar{W}'=(\frac{\bar{c}_1}{c_1}\bar{\mathbf{w}}_{\cdot 1},...,\frac{\bar{c}_n}{c_n}\bar{\mathbf{w}}_{\cdot n})\text{.}
\end{equation}
Since the number of interactions in $\bar{R}$ is much lower than that of $R$, the reliability of similarity relationships in $\bar{W}$ obtained based on $\bar{R}$ should be much lower than that in $W'$ obtained based on $R$.
The column-wise weighting takes into account the difference in reliability due to the number of item interactions.
We use $\bar{W}'$ to quantify the impact of the updated link relationships $\bar{R}$ on item-item similarities.
For negative flip, item pairs with high similarity in $\bar{W}'$ should be weakened in $\tilde{W}$ as follows:
\begin{equation}
\tilde{W}=W'-\bar{W}'\text{.}
\end{equation}
For positive flip, item pairs with high similarity in $\bar{W}'$ should be enhanced in $\tilde{W}$ as follows: 
\begin{equation}
\tilde{W}=W'+\bar{W}'\text{.}
\end{equation}
Finally, we obtain the new prediction matrix as follows: 
\begin{equation}\label{eq:8fam}
\hat{\tilde{R}}=\tilde{R}\tilde{W}\text{.}
\end{equation}
Fig.~\ref{fig:ours} shows the unlearning process of \ours. Note that \ours only requires a model that can provide a mapping matrix, without caring about specific implementation methods.

\subsection{Application Scenarios}

In the previous section, we have introduced two ways to achieve recommendation unlearning for the mapping-based CF methods. The first approach involves correcting only the interaction matrix as~\eqref{eq:gn9m}, while the second approach involves correcting both the interaction matrix and the mapping matrix simultaneously as~\eqref{eq:8fam}.
In fact, in some cases, we can achieve recommendation unlearning by only correcting the mapping matrix:
\begin{equation}
\hat{\tilde{R}}=R\tilde{W}\text{.}
\end{equation}
A User's recommendation results are influenced by both her/his interaction history and the mapping matrix.
\textbf{Correcting a user's interaction records (a row in the interaction matrix) will only affect the user's recommendation results, while correcting the mapping matrix will affect the recommendation results of all users.}
Whether to correct the mapping matrix should consider whether a recommendation unlearning should have an impact on the recommendation results of other users.

In this section, we present some typical scenarios in which \ours can be used to achieve recommendation unlearning, taking into account the requirements for usability and privacy protection.

\subsubsection{Usability}
As shown in TABLE~\ref{tab:case}, we have listed three scenarios for recommendation unlearning based on considerations of system usability: out-of-distribution data, out-of-date data, and attack data.

For out-of-distribution data, interactions that fall outside of a user's true interests often do not exhibit specific patterns and may have a certain degree of randomness, which can affect the calculation of item similarities in the mapping matrix.
Therefore, when users require recommendation unlearning for this purpose, both the interaction matrix and the mapping matrix need to be corrected.

For out-of-date data, some platforms capture users' short-term interests to provide better services, which is crucial.
However, in some cases, users' short-term interests can vanish instantly due to certain events.
The effect of \ours is immediate, which is very beneficial for some special scenarios.
Compared to the examples in TABLE~\ref{tab:case}, there are some examples that may be more urgent.
For example, a person may have just experienced a failed relationship and may not want the recommender systems to recommend memory-evoking items. \ours can make quick changes to this situation.
Since out-of-date data is not harmful data, when users request recommendation unlearning for this purpose, there is no need to correct the mapping matrix, and only the interaction records of the target user need to be corrected.

For attack data, they often exhibit certain patterns that can severely affect the calculation of item similarities in the mapping matrix, thus impacting the overall effectiveness of the recommender system.
When one requests recommendation unlearning for this purpose, the main goal is to mitigate the impact of the attack data on the overall recommendation performance.
Therefore, there is no need to modify users' interaction records, and only the mapping matrix needs to be corrected.

In summary, depending on the specific purpose and requirements of recommendation unlearning, different strategies can be adopted, including correcting both the interaction matrix and the mapping matrix, correcting only the interaction matrix, or correcting only the mapping matrix.
\ours provides the flexibility to address various recommendation unlearning scenarios based on specific needs.

\subsubsection{Privacy}
When users request recommendation unlearning for privacy protection purposes, there are three levels of forgetting:
1) interaction level, which means the user's request is to forget a single interaction in the system;
2) preference level, which means the user's request is to forget all interactions highly related to a specific interaction (for mapping-based CF methods, identifying highly related interactions is straightforward due to the stored similarity relationships); and
3) account level, which means the user's request is to delete all of her/his interactions from the system.

In practice, mapping-based CF methods decouple users' sensitive data (interaction matrix) from model parameters (mapping matrix), so deleting the corresponding interaction records can ensure user privacy without the need to modify the mapping matrix.
Additionally, the data removed for privacy protection purposes is often not harmful data (e.g., out-of-distribution data, attack data), so deleting the model trained with such data might impact the recommendation performance.
However, when recommendation unlearning is requested for privacy protection purposes, the system is required by regulations to delete the model trained with the forgotten data.

We believe that when users request recommendation unlearning for privacy protection, they should be given the option to choose whether to keep the model trained with their interaction data or not, in addition to deleting the interaction records.
This may depend on further legal considerations.
In practice, some recommendation unlearning methods already adopt this approach.
For example, SCIF~\cite{li2023selective} only updates the target user's embedding without considering the impact of forgetting the target data on the embeddings of other users and items.
This means the post-forgetting model retains the knowledge learned from the target data.
In contrast, \ours provides users with the option to choose whether they want the model to be trained with their data or not.

\section{Experiment}
This section validates the superiority of \ours through comprehensive experiments.
Specifically, we first introduce the experimental settings, and then address the following two research questions (RQs):
\begin{itemize}
\item RQ1: How does \ours perform in terms of completeness, utility, and efficiency?
\item RQ2: How does \ours perform in different unlearning scenarios?
\end{itemize}

\subsection{Setting}
In this section, we introduce the experimental datasets, the backbones used to instantiate \ours, the compared baselines, and the hyperparameters of the models.
\subsubsection{Datasets}
\begin{table}[t]
\caption{Dataset Statistics}
\begin{center}
\begin{tabular}{c|cccc}
\hline
\textbf{Dataset} & \textbf{\# User} & \textbf{\# Item} & \textbf{\# Interaction} & \textbf{Density} \\ \hline
ML-1M            & 5741             & 2803             & 824857                  & 5.1\%            \\
Gowalla          & 5992             & 5639             & 281412                  & 0.8\%            \\
Yelp             & 18728            & 15295            & 888154                  & 0.3\%            \\ \hline
\end{tabular}
\label{tab:data}
\end{center}
\end{table}
We conducted experiments on three publicly available datasets:
\begin{itemize}
\item ML-1M\footnote{https://grouplens.org/datasets/movielens/}: A movie rating dataset collected from the MovieLens website~\cite{harper2015movielens}, where an interaction is considered to have occurred if a user rated a movie.
\item Gowalla\footnote{http://snap.stanford.edu/data/loc-gowalla.html}: A location-based social networking website where users share their locations by checking in~\cite{cho2011friendship}, and an interaction is considered to have occurred if a user checked in at a certain location.
\item Yelp\footnote{https://www.yelp.com/dataset}: A user review dataset in which users post reviews on the restaurants they visited. Each user review is considered an interaction.
\end{itemize}

Their statistical information is shown in TABLE~\ref{tab:data}, representing different data sizes and densities.
If a dataset contains ratings, we only retain interactions with ratings greater than or equal to 3.
For each dataset, we adopt the 20-core setting~\cite{he2016vbpr}, which means that we only keep users and items with more than 20 interactions.
We divided the training set, validation set, and test set by the ratio of 7:1:2.

\subsubsection{Backbones}
\ours is a model-agnostic method.
As mentioned earlier, any mapping-based CF method can be used as the backbone for instantiating \ours, and any embedding-based CF method can be first formulated as a mapping-based CF method and then used as the backbone for \ours.
We instantiate \ours using two mapping-based CF methods (SLIM~\cite{ning2011slim} and GF-CF~\cite{shen2021powerful}) and one embedding-based method (MF~\cite{koren2009matrix}), which we refer to as \ours-SLIM, \ours-GF-CF, and \ours-MF, respectively.
The descriptions and details of these three methods have been introduced in Sections~\ref{sec:fga9}.

The selected backbone methods are representative.
On the one hand, GF-CF obtains the mapping matrix through numerical calculations, while SLIM obtains the mapping matrix by optimizing the reconstruction loss.
On the other hand, GF-CF separates compression (or encoding, Fourier transform) and decompression (or decoding, inverse Fourier transform) into two processes, whereas SLIM combines these two processes together.
Therefore, we consider GF-CF and SLIM to represent the majority of mapping-based CF methods.
MF, on the other hand, can be viewed as the initial form of all subsequent embedding-based CF methods.

\subsubsection{Baselines}
We choose retrain from scratch (referred to as \emph{Retrain}) and RecEraser~\cite{chen2022recommendation} as the baseline methods for comparison, as both of them can achieve complete recommendation unlearning.
Regarding SISA~\cite{bourtoule2021machine} and GraphEraser~\cite{bourtoule2021machine}, they can be considered as variants of RecEraser and have been shown to under-perform RecEraser in collaborative filtering tasks, so we do not include them in our comparison.
As far as we know, RecEraser is currently the only open-source recommendation unlearning method.

RecEraser is also a model-agnostic method, but the original paper only provided the implementation for embedding-based CF methods.
Here, we extend RecEraser to mapping-based CF methods.
The process of pretraining, slicing, and training sub-models for RecEraser with mapping-based CF methods is the same as that for embedding-based CF methods, with the only difference lying in the aggregation layer.
In the aggregation layer of embedding-based CF models, it requires aggregating user embeddings and item embeddings from different sub-models.
Since the semantic space of embeddings from different sub-models may differ, a multi-layer perceptron (MLP) is used for feature transformation, which is the motivation behind adding the MLP in the original RecEraser.
However, in the aggregation layer of mapping-based CF models, we need to aggregate the mapping matrices from different sub-models, and each dimension of the mapping matrix has the same meaning, eliminating the need for feature transformation.
Therefore, for mapping-based CF models, we omit the MLP.

We instantiated RecEraser on the mapping-based CF method SLIM and the embedding-based CF method MF for comparison with \ours.
GF-CF cannot be used to instantiate RecEraser because GF-CF is a numerical method and cannot optimize the aggregation layer of RecEraser.

\subsubsection{Hyperparameters}
Since \ours itself does not have hyperparameters, we will only introduce the hyperparameters used for the backbones and baselines.
We determine the optimal hyperparameters by performing grid search and selecting the ones that perform the best on the validation set.
%\begin{itemize}
For SLIM, we set both the L1 regularization coefficient and the L2 regularization coefficient to $1$.
For GF-CF, we set the rank of truncated singular values to $64$.
For MF, we set the dimension of embeddings to $64$ and the regularization coefficient to $0.001$.
For RecEraser, we use MF for pretraining and set the number of shards to $10$. The maximum number of iterations in the slicing process is set to $50$, and the dimension of the attention matrix in the aggregation layer is set to $32$.
%\end{itemize}
The hyperparameters for the sub-models of RecEraser, RecEraser-SLIM and RecEraser-MF, were the same as those used for \ours-SLIM, and \ours-MF.
Additionally, to ensure reproducibility, we fix the random seed to $2024$ for all experiments.

\subsection{General Performance}\label{sec:gaid}
\begin{figure}[b!]
\centerline{\includegraphics[width=0.65\linewidth]{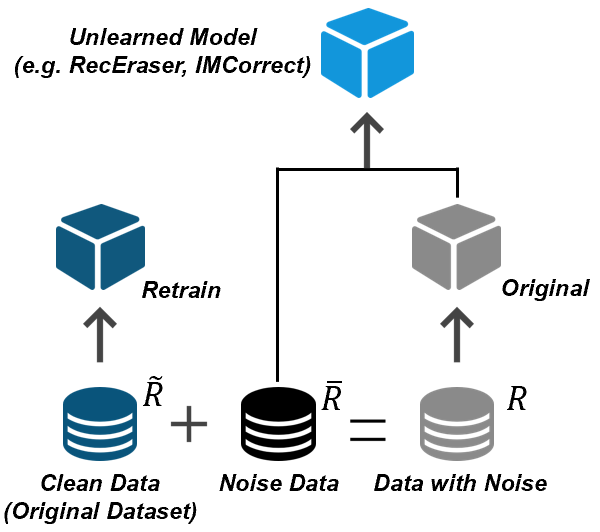}}
\caption{The process of dataset generation.}
\label{fig:gen}
\end{figure}
To evaluate the general performance of a recommendation unlearning method, it should be assessed from three dimensions: completeness, utility, and efficiency.
\subsubsection{Evaluation Methods}
Efficiency can be directly evaluated by comparing the running time of different methods. As for completeness and utility, they can be assessed by comparing the differences in recommendation accuracy between different methods and \emph{Retrain}. However, directly deleting clean interaction data from the training set would lead to a decrease in performance, making it challenging to identify the source of these differences. Completeness and utility are intertwined, making it difficult to effectively evaluate them. For instance, a method may exhibit good utility (small difference from \emph{Retrain}) because it does not achieve high completeness.

To address this issue, three noisy datasets are constructed, and it is assumed that these noisy datasets are the original interaction matrix $R$, with the noisy interactions representing the interactions to be forgotten in the matrix $\bar{R}$. By forgetting these noisy interactions, the original dataset was obtained, corresponding to $\tilde{R}$, the interaction matrix after forgetting. The three methods for constructing noisy datasets are as follows:
\begin{itemize}
\item The ``inert'' dataset, where interactions are randomly added to each user based on the number of interactions in the training set, with proportions ranging from 1\% to 50\%.
\item The ``delete'' dataset, where interactions are randomly deleted from each user based on the number of interactions in the training set, with proportions ranging from 1\% to 50\%.
\item The ``update'' dataset, where interactions are both randomly added and deleted from each user based on the number of interactions in the training set, with proportions ranging from 1\% to 50\%.
\end{itemize}
As shown in Fig.~\ref{fig:gen}, the results of the backbone model running on these noisy datasets are denoted as ``\emph{Original}'', while the results of the backbone model running on the original clean datasets are denoted as ``\emph{Retrain}''. In this setting, forgetting noisy interactions will lead to an improvement in performance because the data in $\bar{R}$ is considered toxic data. This way, completeness and utility are unified. In other words, the better the utility of a method, the better its completeness. This evaluation method is consistent with~\cite{zhang2023recommendation} and~\cite{yuan2023federated}. Recall is used as the evaluation metric to compare the performance of different methods.

\subsubsection{Completeness and Utility Comparison}
\begin{figure*}[htbp]
\centerline{\includegraphics[width=0.99\linewidth]{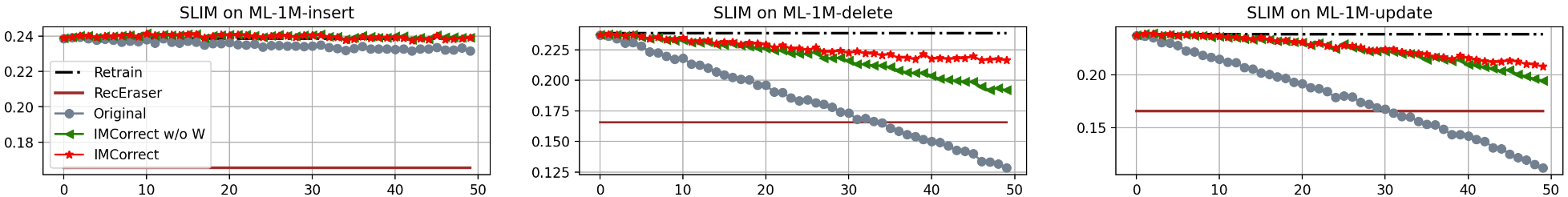}}
\caption{Performance comparison on the ML-1M dataset using SLIM as the backbone.}
\label{fig:ml}
\end{figure*}
\begin{figure*}[htbp]
\centerline{\includegraphics[width=0.99\linewidth]{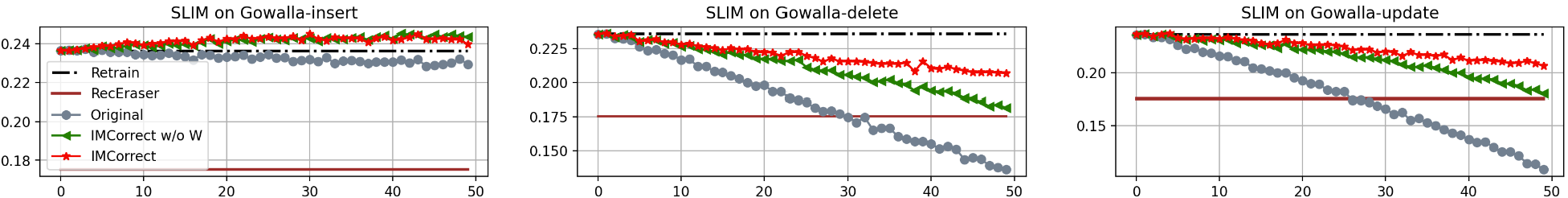}}
\caption{Performance comparison on the Gowalla dataset using SLIM as the backbone.}
\label{fig:go}
\end{figure*}
\begin{figure*}[htbp]
\centerline{\includegraphics[width=0.99\linewidth]{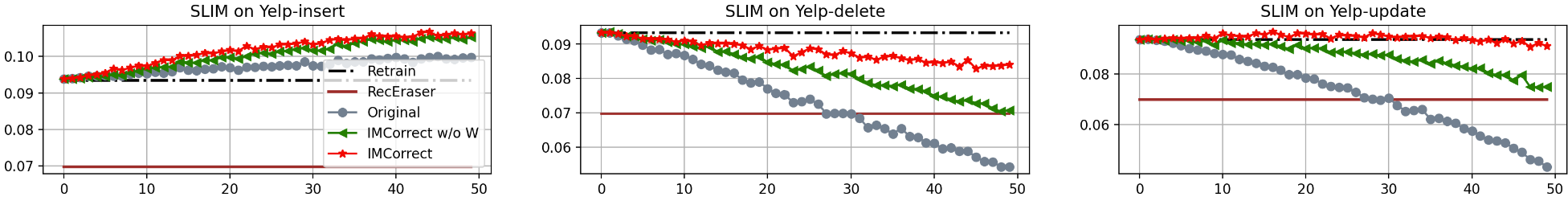}}
\caption{Performance comparison on the Yelp dataset using SLIM as the backbone.}
\label{fig:ye}
\end{figure*}
\begin{figure*}[htbp]
\centerline{\includegraphics[width=0.99\linewidth]{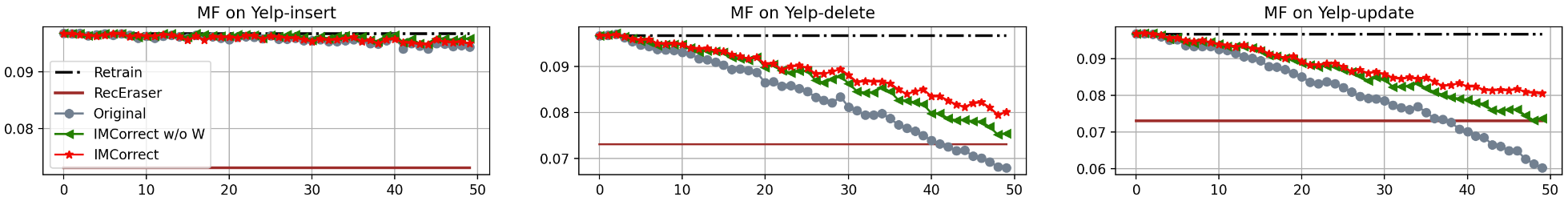}}
\caption{Performance comparison on the Yelp dataset using MF as the backbone.}
\label{fig:mfye}
\end{figure*}
\begin{figure*}[htbp]
\centerline{\includegraphics[width=0.99\linewidth]{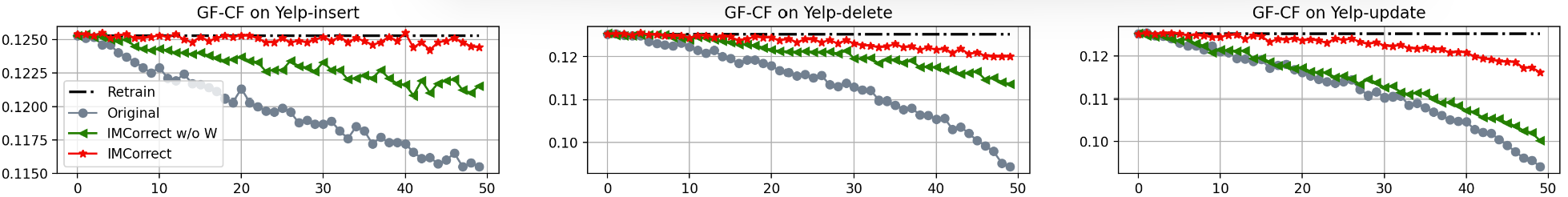}}
\caption{Performance comparison on the Gowalla dataset using GF-CF as the backbone. Note that RecEraser cannot be applied to GF-CF.}
\label{fig:gfye}
\end{figure*}

Due to space limitation, we present the results of implementing different recommendation unlearning methods using SLIM as the backbone on all datasets, and present the results achieved using MF and GF-CF only on the Yelp dataset due to consistent conclusions.
Fig.~\ref{fig:ml}-\ref{fig:gfye} respectively show these results, each with ``insert'', ``delete'', and ``update'' attacks, with the abscissa representing the proportion of attack data to the total training set. From the results, we have the following observations:
\begin{itemize}
\item The performance of \emph{Original} is worse than \emph{Retrain} in all settings. This is because \emph{Original} represents the performance of the backbone model on the data with noise, while \emph{Retrain} represents the performance of the backbone model on the data after forgetting the noises. Since the noisy interactions are toxic data, forgetting them leads to performance improvement.
\item \ours performs better than \ours w/o W (without mapping matrix correction), and \ours w/o W performs better than \emph{Original}. This indicates that efficient recommendation unlearning can be achieved by only correcting the interaction matrix, and correcting the mapping matrix can further enhance completeness and utility.
\item As RecEraser involves retraining, its performance remains consistent for any proportion of noisy data. Although the performance of \ours tends to decline with the increase of noise, even in the most severely attacked scenario where 50\% of interactions are corrupted, \ours (even \ours w/o W) still outperforms RecEraser. RecEraser's performance is even worse than \emph{Original} until the noise proportion in the ``delete'' or ``update'' dataset exceeds approximately 25\%. This suggests that RecEraser's utility is severely compromised due to the partitioning process.
\item Even when random interactions are added for each user with a proportion of 50\% of their total interactions in the ``insert'' dataset, the performance of \emph{Original} remains nearly unchanged. This indicates that the model itself has sufficient robustness against attacks of adding random interactions. As mentioned earlier, mapping-based CF models smooth the interaction matrix to obtain the prediction matrix, acting like a low-pass graph filter. The added random interactions lack specific patterns and represent high-frequency information, so \emph{Original} can achieve decent results.
\item In contrast to the limited impact of the ``insert'' dataset, random deletion of interactions in the ``delete'' dataset has a significant impact on the model performance. This is because the deleted interactions are often the result of users' self-selected choices based on their interests, so they have certain patterns and deleting them will result in a decrease in similarity confidence due to increased sparsity of training data.
\item An interesting phenomenon is observed where, when using SLIM as the backbone, as the noise increases on ``insert'' data, the performance of \ours and \ours w/o W outperforms \textit{Retrain}. On the Yelp dataset, even \textit{Original} can surpass \textit{Retrain}. This phenomenon becomes more pronounced as the sparsity of the dataset increases. A similar phenomenon was observed in~\cite{zhang2023recommendation}. Although the discussion of this phenomenon is not the focus of this paper, we mention it as it also implies that \ours is particularly effective on sparse datasets.
%\item Last but not least, it should be noted that although the performance of \ours decreases on the delete and update datasets, it is noted that even if it is subjected to the most severe attack, where update data accounts for 50\%, it means that half of the interactions in the training dataset are random interactions. The difference in performance between using \ours for forgetting and \emph{Retrain} is still only about 10\%. In addition, the actual amount of forgetting is not significant, so it still has practical value. We present these results to give readers a comprehensive understanding of \ours.
\end{itemize}

\subsubsection{Efficiency Comparison} 
In TABLE~\ref{tab:effciency}, we compare the efficiency of different methods on the ``delete'' dataset with 5\% corrupted data, where the total time of RecEraser is measured by summing the average time of sub-model training and the time of the aggregation layer. Note that we only consider SLIM and MF as the backbones in the comparison because RecEraser cannot be applied to GF-CF. As shown in the results, \ours can be fitted quickly, while the aggregation layer of RecEraser requires training on entire datasets, indicating that \ours can achieve much higher efficiency than \emph{Retrain} and RecEraser.

\subsection{Application Scenario Analysis}

\subsubsection{Usability}
There are three specific scenarios for recommendation unlearning, each corresponding to modifying the interaction matrix or the mapping matrix, or both.
For the out-of-distribution data scenario, it aligns with the noise setting used in Section~\ref{sec:gaid} to validate the general performance of \ours, where specific interaction patterns are often absent.
Since we have already verified the capability of \ours in this scenario, we will not conduct a case study for it.
As for out-of-date data, it essentially involves forgetting specific interactions or preferences, similar to user-initiated forgetting for privacy protection purposes.
The difference lies in that out-of-date data forgetting is often initiated by the system, while forgetting preferences is initiated by users.
Therefore, we only present the results of the case study on attack data.

The experiments were conducted on the ML-1M dataset.
We selected item 2024 and found that it had 55 interactions in the clean original training dataset.
Moreover, we found that when the model was trained using the original dataset, item 2024 appeared in the recommendation lists of all users only 1 time.
Next, we simulated malicious interactions by adding 1000 interactions for item 2024.
At this point, we observed that the occurrences of this item in the recommendation lists increased to 106 times.
Finally, we applied \ours to remove the influence of the malicious interactions, and found that the number of recommendations for item 2024 was restored to 0.
This indicates that \ours can effectively ensure the usability of the model against attacks.

\subsubsection{Privacy}
When users initiate recommendation unlearning for privacy protection purposes, there are three levels: interaction level, preference level, and account level.
Both interaction level and account level can be considered as special cases of preference level.
Therefore, we conducted a case study specifically for the preference level, where we simultaneously correct the interaction matrix and mapping matrix.

The experiment was conducted on the ML-1M dataset.
We selected user 2024 and recorded the recommendations obtained from the model trained on the clean original dataset.
For constructing the forgotten preference, we first randomly select a target item that user 2024 had interacted with, then select 10 most similar items from user 2024's historical interactions, and finally let \ours forget the 11 items. We have repeated the above process 100 times and compared the recommendation lists before and after forgetting. In the results, the average similarity between the recommended items and the target item is 0.0136 before forgetting and 0.0063 after forgetting, i.e., the recommended items are much less similar to the target items after forgetting, indicating the effectiveness of \ours in forgetting user preferences (i.e., protecting user privacy) from the recommendation models.

\begin{table}[t!]
\caption{Running Time Comparison (``m'' Stands for Minute and ``s'' Stands for Second)}
\label{tab:effciency}
\begin{center}
\begin{tabular}{c|ccc}
\hline
\textbf{Model}                                                                                  & \textbf{ML-1M}              & \textbf{Gowalla}            & \textbf{Yelp}                \\ \hline
Retrain-SLIM                                                                                    & 13.0m                       & 5.1m                        & 8.5m                        \\\hline

\multirow{2}{*}{\begin{tabular}[c]{@{}c@{}}RecEraser-SLIM\\ (sub-model+aggregate)\end{tabular}} & \multirow{2}{*}{2.1m+11.7m} & \multirow{2}{*}{1.2m+6.6m} & \multirow{2}{*}{2.3m+27.4m}  \\                                                                                                &                             &                             &                              \\ \hline
{\bf \ours-SLIM}                                                                                        & 30s                         & 15s                          & 1.6m                          \\\hline\hline
Retrain-MF                                                                                      & 24.7m                       & 16.8m                       & 32.7m                        \\\hline

\multirow{2}{*}{\begin{tabular}[c]{@{}c@{}}RecEraser-MF\\ (sub-model+aggregate)\end{tabular}}   & \multirow{2}{*}{5.4m+12.1m} & \multirow{2}{*}{3.2m+7.1m} & \multirow{2}{*}{8.4m+32.2m} \\                                                                                             &                             &                             &                              \\ \hline
{\bf \ours-MF}                                                                                          & 2.2m                         & 1.9m                         & 5.6m                          \\\hline
\end{tabular}
\label{tsab2}
\end{center}
\end{table}
\section{Conclusions}
%Recommendation unlearning, which involves forgetting specific data and models, can protect users' right to be forgotten and improve system usability.
%To ensure the efficiency of forgetting, a recommendation unlearning method often needs to strike a balance between completeness or utility and efficiency.
%Based on our analysis of the essential meaning of the mapping matrix in mapping-based CF models, 

In this paper, we propose a model-agnostic recommendation unlearning method named \ours.
\ours achieves high efficiency in forgetting by only correcting the interaction matrix, and can further achieve high completeness and utility by correcting the mapping matrix.
Detailed experiments and case studies show that \ours can achieve high completeness, utility, and efficiency, and can be applied to many real-world recommendation unlearning scenarios.
%paving the way for future research in this important area. 
This paper primarily focuses on validating its decremental learning capability. It should be noted that \ours has the ability for incremental learning, we will further explore the performance of \ours in incremental recommendation scenarios in future works.

\bibliographystyle{IEEEtran}
\bibliography{reference}

\end{document}